# Objective task-based evaluation of artificial intelligence-based medical imaging methods: Framework, strategies and role of the physician

Abhinav K. Jha[1,2,3,*], Kyle J. Myers[4], Nancy A. Obuchowski[5], Ziping Liu[1], Md Ashequr Rahman[1], Babak Saboury[6], Arman Rahmim[7], Barry A. Siegel[2,3]

[1]Department of Biomedical Engineering, Washington University in St. Louis, MO, USA
[2]Mallinckrodt Institute of Radiology, Washington University in St. Louis, MO, USA
[3]Alvin J. Siteman Cancer Center, Washington University in St. Louis, MO, USA
[4]Center for Devices and Radiological Health, Food and Drug Administration, Silver Spring, MD, USA
[5]Quantitative Health Sciences, Cleveland Clinic, Cleveland, OH, USA
[6]Department of Radiology and Imaging Sciences, National Institutes of Health, Bethesda, MD, USA
[7]Departments of Radiology and Physics, University of British Columbia, BC, Canada
*Corresponding author: a.jha@wustl.edu

**Synopsis**: Artificial intelligence (AI)-based methods are showing promise in multiple medical-imaging applications. Thus, there is substantial interest in clinical translation of these methods, requiring in turn, that they be evaluated rigorously. In this paper, our goal is to lay out a framework for objective task-based evaluation of AI methods. We also provide a list of tools available in the literature to conduct this evaluation. Further, we outline the important role of physicians in conducting these evaluation studies. The examples in this paper will be proposed in the context of PET with a focus on neural-network-based methods. However, the framework is also applicable to evaluate other medical-imaging modalities and other types of AI methods.

**Key points**:

1. Important need for strategies for rigorous objective evaluation of artificial intelligence (AI)-based methods on clinical task.
2. We lay out a framework for objective task-based evaluation of AI-based methods.
3. Methods to conduct such evaluations, specifically in the context of PET, are presented.
4. The role of physicians in conducting such evaluation is presented.
5. Examples of applying the framework to evaluate hypothetical AI-based methods for PET data acquisition and reconstruction are presented.
6. Future areas of research on task-based evaluation are outlined.

**Disclosure statement**: Nancy Obuchowski is a statistician for the Quantitative Imaging Biomarkers Alliance (QIBA).

# A. Introduction

Artificial intelligence (AI)-based methods for medical imaging, and more specifically positron emission tomography (PET), hold exciting promise in multiple stages of the imaging-technology-development lifecycle ranging from data acquisition to image reconstruction, image processing, clinical assessment, and towards clinical decision support[1-3]. Thus, there is substantial interest in the clinical translation of these methods. While AI encompasses a broad range of methods, in this manuscript, we specifically refers to methods that are based on artificial neural networks, such as deep-learning approaches. These methods have been showing the most promise in medical imaging over the past few years.

For translation of any technology, the need for rigorous evaluation is well recognized[4]. The need for such evaluation is even more critical for AI methods. These methods are generally not programmed with user-defined rules and instead learn rules by analysis of training data. For example, a neural network-based method for tumor segmentation is not provided any guidance or rules on the defining traits of a lesion boundary (e.g. difference in tumor and background intensities). Instead, the method is trained on example images (referred to as training images) where a delineation of the tumor boundaries is given, and the method learns the rules to segment these tumors. These rules are implicit and not necessarily interpretable (although there are several efforts towards interpretability), thereby making the output of these methods unpredictable and not explainable. This can have major implications. One implication is the inability to detect failure of the method. Evaluations are required to identify such failures, assess the reliability of the methods, and thus provide guidance on clinical applicability. Next, since the rules learned by the AI methods are derived from a training dataset, their performance with new unseen datasets, such as from a different patient population, or acquired with a different imaging protocol, may be sub-par (issue of generalizability)[5,6]. Evaluations would qualify this generalizability. A related issue is that AI-based methods learn spurious correlations from training data that may be unrelated to disease state[7-9]. A recent study, rather alarmingly, showed that even highly publicized AI-based systems for COVID-19 detection from chest radiographs relied on confounding factors rather than medical pathology[9]. Evaluations can help identify such issues. Further, AI methods are being explored in new areas such as fully automating medical-imaging applications that currently require some level of manual intervention[10] and in making critical therapeutic decisions[11]. The high costs associated with inaccuracy in these decisions make rigorous evaluation even more critical. Consequently, there have been multiple calls for rigorous evaluation of AI methods[12-15].

## Concept of Clinical Task in Medical Imaging:

In medical imaging, images are acquired for specific clinical tasks such as detection, quantification, or a combination. For example, an oncological PET image in a patient showing symptoms of lung cancer may be acquired for the task of tumor detection, tumor-tracer-uptake quantification, or both. Thus, it is widely recognized that evaluation of medical-imaging methods should be performed on the clinical task that is required from the images. However, current evaluations of AI methods are typically task agnostic. For example, AI-based reconstruction and post-processing methods are typically being evaluated using figures of merit (FoMs) such as root mean square error (RMSE) that measure fidelity between the estimated image and a certain reference standard (see accompanying [Joyita-chapter-xx] in this issue for a listing of some of these FoMs). However, studies are showing that these FoMs may provide misleading interpretations[16-20]. For example, evaluation of an AI-based reconstruction approach for whole-body FDG-PET using these FoMs suggested that the method was yielding similar performance as a conventional approach. However, evaluation on the task of lesion detection revealed that the method was yielding false-negatives due to blurring or missing lesions, and false positives due to pseudo-low-uptake patterns[16]. Another study quantitatively compared task-based evaluation with evaluation using fidelity-based FoMs for an AI-based denoising method for low-dose cardiac SPECT images[17]. In that study, the fidelity-based FoMs suggested that the AI-based denoising led to improved performance. However, on the task of detecting cardiac perfusion defects, the clinical task for which these images are acquired, the performance with AI-based denoising method was almost equivalent, if not worse, compared to without conducting the denoising. Similar findings have been observed in subsequent studies[18-20]. Similarly, AI-based methods for PET image segmentation are commonly evaluated using FoMs such as Dice scores, which quantify some measure of distance between the estimated segmentation and a reference standard (e.g. manual segmentation). However, again, these FoMs do not directly correlate with the task required from the image, such as estimation of quantitative features[21] or PET-based radiotherapy planning[22]. In summary, the current task-agnostic approaches to evaluating AI methods have limitations. Thus, there is an important and timely need for strategies to conduct task-based evaluation of these methods.

The literature on objective assessment of image quality (OAIQ) has proposed multiple techniques to achieve the goal of objective task-based evaluation[23-26]. Further, OAIQ techniques have been widely applied to evaluate imaging methods[27], including in PET[28-30]. However, designing OAIQ studies to evaluate AI methods requires special considerations given the unique working principles of these methods in terms of learning rules from training data. Our goal in this work is to propose an OAIQ-based framework that is fine tuned for evaluating AI methods. While the framework is general, it is presented in the context of PET. The key criteria when designing an OAIQ study to evaluate AI methods are summarized in Table 2.

A few remarks: the framework presented here is readily extendible to SPECT and a number of other imaging modalities. In fact, recent studies have investigated evaluating AI-based SPECT methods using this framework[17,31]. Further, we focus on evaluation of AI methods that are intermediate to performing the eventual clinical task. These include methods for optimizing imaging system instrumentation and acquisition protocols, image reconstruction, any kind of image processing, and image analysis. Finally, while we propose this framework in the context of evaluating neural network-based approaches, the framework could also be applied to evaluate other categories of AI methods.

## B. Framework for objective task-based assessment

A rigorous and structured evaluation of imaging methods using the OAIQ framework requires four essential steps[27]:

1. Specification of the task
2. Defining the patient population and the imaging process
3. Method to extract task-specific information
4. Figure of merit

Below, we discuss each of these components in the context of evaluation of AI-based methods for PET. A well-designed evaluation study should yield a claim that quantitatively defines the performance of the method[32,33]. Following this four-component framework would yield such a claim, as we will see in the evaluation of hypothetical AI-based methods in a following section.

## OAIQ terminology

We first introduce some OAIQ terminology with reference to clinical imaging. As per this terminology, the biological property of the patient that is measured during the imaging process is referred to as the object. In PET, this would be the distribution of the tracer within the patient. The object consists of signal and background. The signal typically refers to the abnormality in the patient, while the background refers to regions in the patient in the absence of the signal. For example, consider a patient showing symptoms of cancer who has been injected with, say, the 18F-FDG tracer. The concentration of this tracer in the cancer cells and in the normal tissues would be referred to as the signal and background, respectively. The measured data is referred to as the image. In PET, this can be the sinogram, but often also refers to the reconstructed image.

We also introduce the imaging equation, which will assist with illustrating the framework. Consider a PET system described by an operator $\mathcal{H}$ that is imaging a radiotracer distribution, denoted by a vector $f$, within a patient. This tracer distribution is a continuous function. The imaging process leads to measured projection data $g$, a finite-dimensional vector. Further, during imaging, noise, denoted by the vector $n$, is introduced. The imaging equation is then given by

$$g = \mathcal{H}f + n \qquad (1)$$

Often, the projection data is reconstructed, yielding a reconstructed image denoted by $\hat{f}$. In these cases, to describe the transformation from the object $f$ to the image $\hat{f}$, the reconstruction operator also needs to be incorporated into the imaging equation.

With this background, we provide the four-component evaluation framework below.

## B.1 Specification of the task

PET images are acquired for multiple clinical tasks. Typically, these tasks can be divided into three broad categories (Fig. 1). They are briefly described with corresponding examples.

### B.1.1 Classification task

A task that results in a decision-maker/observer deciding in favor of a particular hypothesis from a finite hypothesis space is termed a classification task. If the space contains only two hypotheses (yes/no decision), then the task is termed as binary classification. Examples include lung-tumor detection using an FDG-PET scan, cardiac-defect detection using myocardial perfusion PET and differential diagnosis of Alzheimer's disease from frontotemporal dementia using FDG-PET in patients with dementia. This binary-classification task is also termed as a signal-detection task where presence of a signal (increased uptake in lung nodule, reduced blood flow in myocardium, decreased uptake in brain regions) typically corresponds to abnormality.

Depending on the information available about the to-be-detected signal and the background, the detection task can be classified into different categories. The signal can be known exactly (SKE) or statistically (SKS), and likewise, the background can be known exactly (BKE) or statistically (BKS). Note that an SKE or BKE task does not imply that the signal or background is flat. Instead, the implication is that the tracer uptake at each point within the signal or background, whether homogeneous or heterogeneous, is known. In contrast, for an SKS or BKS task, only a statistical description of this tracer uptake would be available. In clinical settings, both the signal and the background are known only statistically. However, tasks that are SKE or BKE can help get insights on imaging system performance[34]. Another point to emphasize here is that even in a task that is both SKE and BKE, the image will only be known statistically, since imaging systems introduce noise (Eq. 1).

In some cases, the task can be posed as one where the signal properties are variable, but in each image presented to the observer, the observer knows the signal properties, although they do not know whether the signal is present. This is referred to as a signal-known-exactly-but-variable (SKEV) task. For example, in each image presented to the observer, the lesion location is different, but known to the observer. The observer only has to decide if the lesion is present at that location.

When evaluating imaging methods, realistic modeling of variability in the patient population is important. Thus, tasks that model this variability, which include SKS/BKS and SKEV/BKS tasks should preferably be considered. The data collected for these tasks should be representative of patient population, as we further describe in the next section. However, in certain settings, an SKE/BKS task may be considered to evaluate the performance of the method for specific signal sizes or amplitudes or study the sensitivity of the method to these signal properties.

**B.1.2 Estimation/Quantification task**

The goal in estimation/quantification tasks is to measure some numerical or statistical feature of the object that has been imaged. In PET and SPECT, this includes quantifying features such as tracer uptake within a certain region of interest (ROI)[35-38], and volumetric parameters such as metabolic tumor volume (MTV) and total lesion glycolysis (TLG)[10]. Additionally, in tracer-kinetic modeling, parameters that describe the physiological characteristics of an ROI, such as blood flow and receptor binding potential, are estimated[39]. Further, quantitative PET (and SPECT) are being actively explored for targeted radionuclide therapy dosimetry[40]. In summary, a broad range of quantification tasks are performed using radionuclide imaging.

When evaluating AI methods, the task should be posed as one where the features of both the signal (e.g. tumor) and the background are varying, preferably realistically as in clinical settings. This is equivalent to an SKS/BKS setting in the context of estimation tasks.

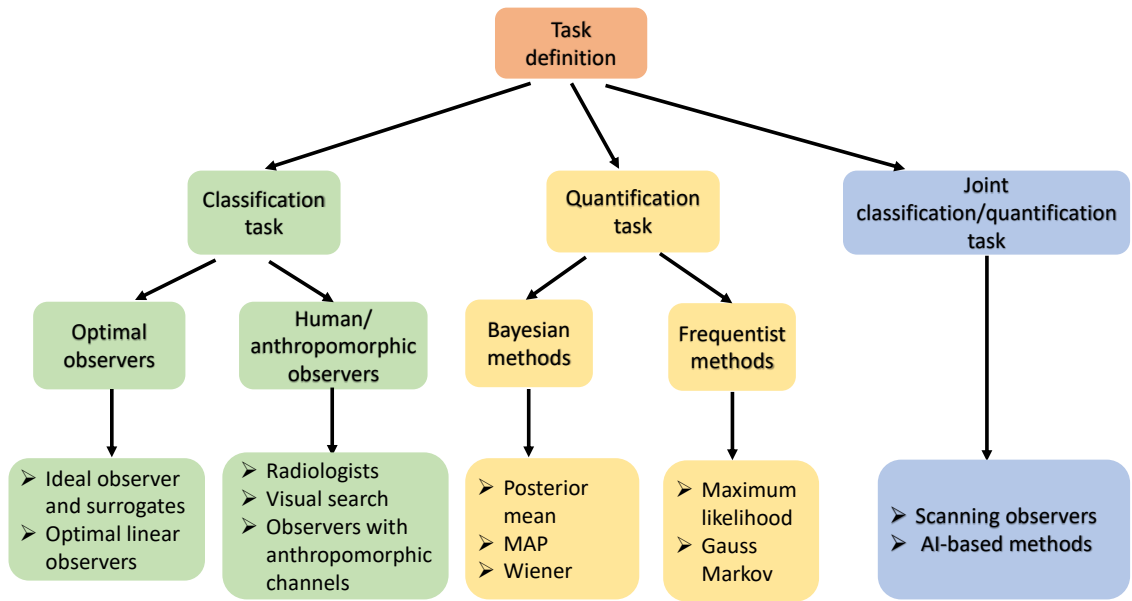

Fig. 1: A taxonomy of the clinical tasks in PET, and the methods to perform these tasks.

**B.1.3 Joint classification and quantification task**

Often, the clinical task in PET involves both classification and quantification of some features of an abnormality. For example, to quantify MTV of the primary lung tumor in a patient with lung cancer from an FDG-PET image, the primary tumor has to be first detected. These are referred to as joint classification and quantification tasks. Again, when evaluating AI methods, the detection and quantification components of the task should account for patient variability.

## B.2 Define Patient Population and Imaging Process

Rigorous image quality assessment should model the physical and statistical properties of the objects (i.e. the patients) being imaged through this system. Thus, the patient population selected for evaluation should be representative of populations seen in clinical practice. Factors that would impact method performance, which could include sex, age, ethnicity, and disease prevalence, should be accounted in the study. Further, depending on the method, the image-acquisition protocol, including the system(s) over which the patient images were acquired, whether the study was single/multi center, and other acquisition settings, should be specified. This would also help quantify the generalizability of the AI method to different populations and/or different image-acquisition protocols. For example, multi-center evaluation will provide more confidence about generalizability compared to a single-center evaluation.

An OAIQ study typically requires some level of statistical description of the image data to extract task-specific information. We see from Eq. 1 that defining the patient population and the imaging process can yield such a description. This description can be obtained through analytical approaches, clinical studies, and realistic simulations. Of these three approaches, clinical studies typically provide the most confidence for clinical translation. However, conducting these studies presents several challenges (high costs, patient risks, time consuming) and thus may not be feasible for each AI method. Further, clinical studies may not have ground truth available for conducting the evaluation. Clinically realistic simulations, as is conducted in virtual-clinical-trials (VCTs)[41,42], provide a mechanism to address these issues and identify promising methods for further clinical evaluation. In the next sub-section, we present the tools for conducting these simulations for PET. Similar tools are available in other modalities

### B.2.1 Tools to conduct clinically realistic simulations

In clinically realistic simulations, image data corresponding to a clinically realistic digital model of the patient population is generated using software that models the imaging scanners. These simulations provide several advantages such as known ground truth, the ability to simulate patient variability and imaging physics, generation of numerous images relatively quickly and inexpensively, minimal risks (e.g., no radiation dose) and no patient discomfort. These images can then be used to conduct an OAIQ study. In the rest of this subsection, we describe the tools available to simulate patient populations, simulate PET systems and factors to consider while reconstructing the images.

### B.2.2 Description of the patient population:
To ensure clinical relevance in simulation-based evaluations, accurate modeling of *in vivo* anatomical and physiological properties of the patient and variability in patient population is essential. Anthropomorphic digital phantoms can be used to achieve this goal. Kainz et al. provide a library of available anthropomorphic phantoms[43]. A widely used anthropomorphic phantom is the extended cardiac and torso (XCAT) phantom[44], which can generate digital phantom populations with variable anatomies, including scaling the size of the body or specific organs, modeling normal and diseased patients, and modeling cardiac and respiratory motions. Additionally, hybrid phantoms can be used where simulated signals are added to real clinical images[45,46] to ensure background realism. Another option is stochastic approaches that estimate parameters that describe the object distribution directly from patient images[47]. Generative adversarial network (GAN)-based approaches have also been explored to simulate phantom populations[48].

Next, realistic modeling of lesions (the signal) is important. Assuming simplistic models such as spherically-shaped lesions can provide misleading interpretations. To develop realistic models, we can obtain a distribution of the parameters of lesions, such as size, shape, and signal-to-background uptake ratio, from clinical images. These distributions could be sampled to generate lesions. Leung et al. used such an approach to simulate the primary tumor in patients with lung cancer[45]. Additionally, it is preferable to account for heterogeneity within the lesion, such as intra-tumor heterogeneity in oncological PET. We can define this heterogeneity directly from the manually delineated tumor on PET images[49]. However, this may limit the number of simulated lesion types. An alternative is sampling heterogeneity descriptors from a distribution[46].

In simulation studies, the phantom must be generated at a higher resolution than the reconstructed image. This simulates the continuous radio-tracer distribution, is clinically realistic, and has been shown to yield realistic reconstructed images[50]. Also, imaging systems are typically unable to measure certain components of the object function. For example, a low-resolution imaging system, as PET and SPECT systems often are, will not pass the high frequency components in spatial domain. Similarly, a system that records the measurements as detector pixel values will inherently transform a continuous object to a discrete vector (Eq. 1), leading to loss of high-frequency components[12,26,51]. These components of the object that the imaging system is unable to measure are referred to as null functions[12,26]. Almost all biomedical imaging systems have a null space. Simulating the phantom at a higher resolution enables investigating the impact of null space of the imaging system on task performance.

Table 1: Comparison of PET simulation software tools.

| Software | GATE | SimSET | PeneloPET | PETSTEP | SMART PET | ASIM |
|---|---|---|---|---|---|---|
| Category | MC | MC | MC | Analytical | Analytical | Analytical |
| Computational expense | High | High | High | Low | Low | Low |
| PSF modeling | Yes | Yes | Yes | Yes | Yes | Yes |
| Attenuation modeling | Yes | Yes | Yes | Yes | Yes | No |
| Scatter modeling | Yes | Yes | Yes | Yes | Yes | No |
| Random modeling | Yes | Yes | Yes | Yes | Yes | No |

**B.2.3 Modeling PET Instrumentation:** The next step is to generate PET measurements (in projection data or list-mode format) for the digital-phantom population. Realistic simulations should accurately model clinical PET-scanner configurations, including transaxial and axial field of view, detector crystal pitch, and other acquisition parameters. Additionally, modeling photon-propagation and detector physics including attenuation, scatter, random coincidences, noise, and resolution degradation due to positron range, photon non-collinearity, and inter-crystal penetration increases the realism of simulations.

Multiple PET simulation tools have been developed for modeling PET systems and can be divided into two major categories, namely Monte Carlo (MC) and analytical simulations (Table 1). The MC simulation tools, such as GATE[52,53], SimSET[39], and PeneloPET[54] accurately model imaging physics, but are computationally expensive. Task-based evaluation may require generating hundreds or even thousands of images, which may limit the feasibility of using these tools. An alternative is analytical simulation software that model only the major image-degrading processes but are computationally fast[45,46,55,56].

**B.2.4 Modeling image reconstruction and post-processing protocols**: Multiple AI methods, such as those for post processing and segmentation, are designed to operate on reconstructed images. To evaluate these methods, reconstructed images must be simulated. To obtain these images from simulated sinograms, clinical protocols should be followed. For example, when reconstructing using ordered subsets expectation maximization (OSEM)-based methods, the number of iterations and subsets should be similar to those in clinical protocols. Further, depending on the clinical protocol, the reconstruction process may have to compensate for image-degrading processes.

## B.3 Process to extract task-specific information

The third step in the OAIQ-evaluation framework is defining the process to extract task-specific information from the acquired image. The entity that extracts this information is termed as an observer. In general, any observer can be described by an operator $\mathcal{T}$ that operates on some projection data $g$ or the reconstructed image $\hat{f}$ to estimate a statistic. For a detection task, this statistic can be compared to a threshold to decide about the presence of the signal (Fig. 2). For a quantification task, this statistic could be the feature that was intended to be estimated (Fig. 3). The process to extract this information for various tasks is summarized below.

### B.3.1 Classification task

For a classification task, the term "observer" conjures the image of a trained radiologist. The radiologist belongs to the class of human observers. However, human-observer studies can be expensive, time-consuming, and tedious. Further, often OAIQ studies need to be performed over hundreds or even thousands of images and several different configurations, exacerbating these issues. As an alternative, numerical observers, referred to as model observers[57,58], have been proposed. These observers are mathematical algorithms that extract a test statistic from the projection data or the reconstructed image. This test statistic can then be compared to a certain threshold for decision making on the detection task.

Of these model observers, the observer that utilizes all statistical information available regarding the task to maximize task performance is referred to as the ideal observer (IO). The IO provides the best possible performance for the detection task. Computation of the IO for SKS/BKS tasks can be challenging, but strategies have been developed to address this issue. Park et al.[59] and Zhou et al.[60] proposed strategies to compute the IO test statistic for parametric object models. Clarkson et al. proposed and applied Fisher information-based surrogates to the IO[61,62]. However, the IO can still be challenging to compute in clinically realistic settings.

An alternative to IO is using optimal *linear* observers. These have been developed for SKS/BKS and SKEV/BKS tasks[63-65]. For example, Li et al. proposed and theoretically demonstrated the optimality of a multi-template linear discriminant observer for SKS tasks[65]. However, computing these observers can still be computationally challenging due to the large dimension of the image data $g$. The use of channels provides a mechanism to address this issue. Briefly, an operator, columns of which correspond to a small number of channels, are applied to the image data $g$, yielding a much lower-dimensional vector. To approximate the performance of the IO, a class of channels referred to as efficient channels must be used.

Some commonly used linear observers in OAIQ studies are the Hotelling observer and non-prewhitening matched filter, but these should be used with caution as they make certain assumptions that may not hold for SKS/BKS tasks[66].

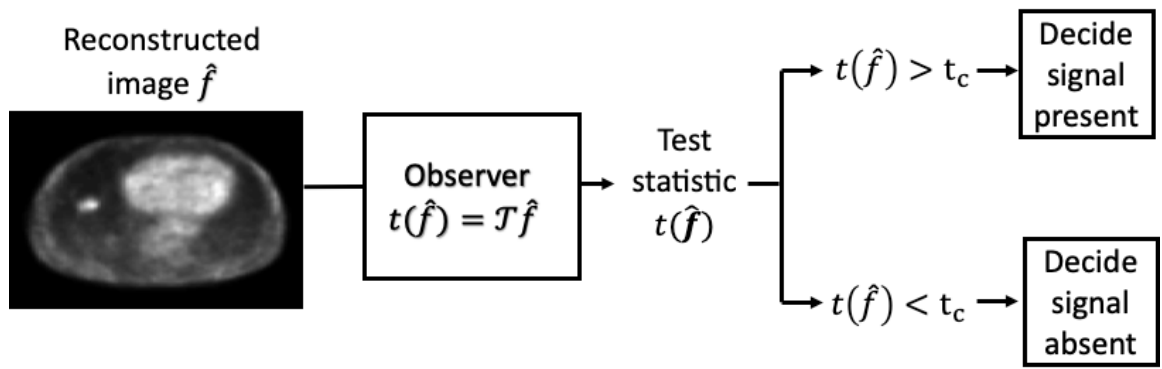

Fig. 2: A schematic demonstrating the process that an observer follows to decide on the presence of a signal when interpreting a reconstructed image. In the figure, $t_c$ denotes the decision threshold.

When the goal of the AI method is optimizing system-hardware design or acquisition protocol, performing the detection task directly on projection data is recommended. This enables the OAIQ study to purely evaluate system performance, and be agnostic to the choice of the reconstruction method. Further, for system optimization, use of an optimal observer, such as the IO, is recommended.

Many AI methods are developed for applications where the detection task is performed on a reconstructed image that is displayed for visual interpretation. These include applications such as reconstruction and image processing. In this case, psychophysical studies with trained human observers, such as radiologists, are the gold-standard. Standardized methods are available to conduct studies to evaluate performance on detection tasks[67]. These include the two alternative forced choice (2-AFC) test and the rating-scale-based experiments. Due to the inherent variability in readers' diagnostic accuracy, on account of differences in training, experience, and perceptual and cognitive abilities, studies to evaluate performance on detection tasks usually involve several trained readers to capture the spectrum of performances. For this purpose, a rich literature on multi-reader multi-case (MRMC) study design and analysis methods exists[68,69]. A key criterion while designing these studies within an OAIQ framework is to evaluate performance on diagnostic tasks, and not on factors such as goodness of images, ease of interpretation, or fidelity-based similarity to some benchmark.

When human-observer studies are not feasible, model observers that mimic the human visual system, also referred to as anthropomorphic observers, are an excellent substitute. Previous experiments in human vision have shown that humans process data through the use of frequency-selective channels. By incorporating such channels in the observer model, studies have shown that model observers can mimic human-observer performance[34,70,71]. These channels are referred to as anthropomorphic channels. Additionally, several other anthropomorphic observers have been proposed in literature[58], including for SKS/BKS tasks[72,73]. Recently, deep-learning-based observers have shown promise in predicting human performance in SKS/BKS tasks[74]. When using an AI-based observer to evaluate an AI-based imaging method, it should be ensured that they are both independently designed and validated.

### B.3.2 Estimation/Quantification task

In an OAIQ study, the goal of the estimation task is to use image data to estimate parameters that define the object[26]. In contrast, in areas such as texture analysis or computer-aided diagnosis, the typical goal is estimating parameters that define the image. For example, when estimating MTV or intra-tumor heterogeneity, an OAIQ study will define these as the properties of the tumor and then estimate those from the image by incorporating the transformation of the imaging system and the noise (Eq. 1). In contrast, in most texture-analysis studies, the tumor-image heterogeneity, and not tumor heterogeneity is estimated. Note that such heterogeneity may simply be introduced by noise or image-degrading processes in PET.

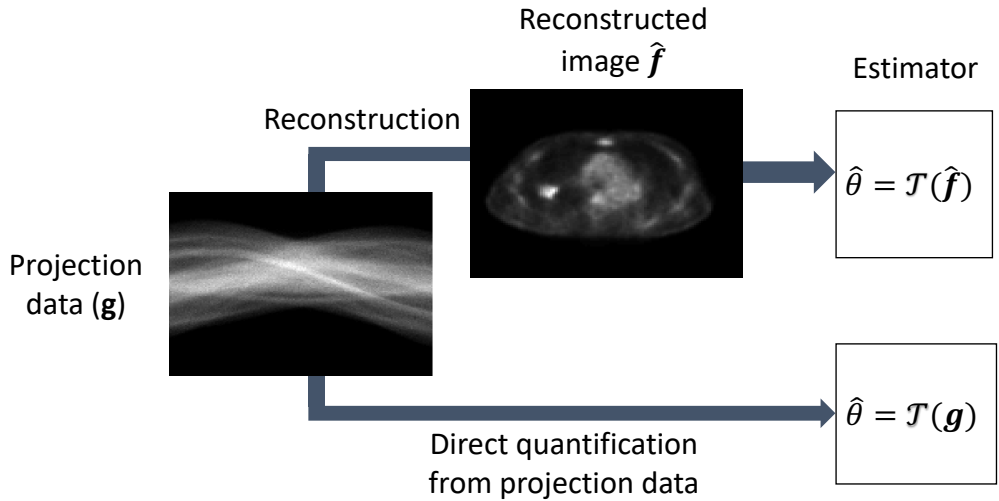

Fig. 3: Schematic demonstrating the process to estimate a parameter of interest from projections and reconstructed images.

Another important concept is that of estimability. The null space of the imaging system impacts the estimation of parameters of the object[75,76]. A parameter that can be estimated accurately from the measured image data for all true values of the parameter in the presence of these null functions is referred to as an estimable parameter.

Consider an object described by some parameters, of which we are interested in estimating the parameter denoted by $\theta$. The estimation task can be posed as an operator that maps from $g$ to an estimate of $\theta$, denoted by $\hat{\theta}$. These tasks are typically performed by first reconstructing the PET image over a voxelized grid, and then estimating the parameter of interest from that reconstructed image (Fig. 3). A less common but highly effective approach is to directly estimate the unknown parameters from the projection data[38,75,77]. Similar to detection tasks, when the goal is system optimization, quantify the parameters directly from projection data is recommended. However, when evaluating AI methods for reconstruction or post-reconstruction operations, the quantification task should be performed on reconstructed images.

Multiple quantification methods have been developed[27,78], which can be broadly categorized into two categories, frequentist and Bayesian, as described below:

**B.3.2.1 Frequentist methods**: In these methods, the parameter $\theta$ for a given patient is assumed fixed. A widely used method in this category is maximum-likelihood (ML) estimation. This estimator has multiple desirable properties. In particular, the estimator achieves the lowest bound of variance for any unbiased estimator (Cramer-Rao lower bound), if this bound can be achieved. An estimator that achieves this bound is referred to as efficient, so the ML estimator is efficient if an efficient estimator exists.

**B.3.2.2 Bayesian methods**: In Bayesian methods, for a given patient, $\theta$ is assumed to be a random variable and a prior PDF on $\theta$ is assumed. The development of these methods starts with defining a cost function. When the cost function is the ensemble mean-squared error (EMSE) between true and estimated value of the parameter of interest, the posterior mean of $\theta$ minimizes that cost function[27]. The posterior mean also minimizes binary cross-entropy loss[79], and in fact, any symmetric convex upward cost function between the measured and the true values[27]. Similarly, the maximum-a-posteriori estimator minimizes the cost function that is constant everywhere other than when the difference between the measured and true values is small, where it is zero.

The estimators mentioned above are typically non-linear in data. Linear estimators have also been proposed. These estimators map the data to the estimate using a linear transformation. Among all linear estimators, Weiner estimator achieves the lowest EMSE and has been used for optimizing imaging systems[80]. A scanning linear estimator has been proposed[81] that has exhibited reliable performance in tasks such as estimating signal location.

### B.3.3 Joint classification and quantification task

Research on developing observers for the joint classification/estimation task is relatively in infancy. Some proposed observers include the channelized joint observer[82-84] and the channelized scanning linear observer[85]. AI-based approaches are also showing promise in this area[86].

## B.4 Figures of merits (FoMs) to measure task performance

FoMs provide a quantitative measure of the performance of the imaging method on a certain task. Below we summarize the FoMs for the tasks discussed above.

### B.4.1 Classification task

Consider a binary classification task where the object being imaged (the patient) belongs to one of two classes: signal present vs. signal absent. In this task, the true positive fraction (TPF) and false positive fraction (FPF) are defined as the proportion of correctly defined signal-present decisions and incorrectly defined signal-absent decisions (Fig. 4a). Commonly, the TPF is referred to as sensitivity and 1–FPF is referred to as specificity. However, the sensitivity and specificity are computed based on a specified decision threshold, and thus can vary based on the strictness of this threshold. A complete characterization of observer performance can be obtained by plotting the TPF and FPF at different threshold values. This yields a receiver operating characteristic (ROC) curve (Fig. 4b). This curve provides the most comprehensive description of detection-task performance[87]. Further, the area under the ROC curve (AUC) provides a summary FoM to evaluate detection-task performance. The observer yielding a higher AUC is more accurate on the corresponding detection task. For tasks that involve classifying the object into more than two classes, multi-class ROC analysis techniques have been proposed[88-92].

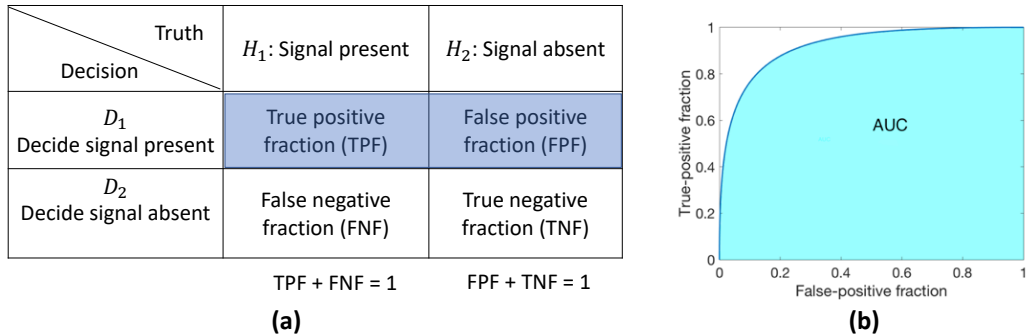

Fig. 4: Schematics illustrating the concepts of (a) TPF and FPF and (b) ROC curve and AUC

**B.4.2 Quantification task**

In quantification tasks, criteria to evaluate performance can include accuracy, precision, and overall reliability. We describe these below, along with the corresponding FoMs to evaluate performance on these criteria:

a. The accuracy of a measurement is the closeness between the true and average measured values, and is commonly quantified by measurement bias. However, the bias is often a function of the true value. A bias profile, which shows the bias as a function of the true value, provides a more complete measure of performance[93,94]. To account for variability in the parameter of interest across populations, ensemble bias, which is the bias averaged over the distribution of the true values, is a more appropriate choice as a summary FoM for accuracy.
b. The precision of an estimate is the repeatability of the estimate over multiple noise realizations of the projection data. This precision is quantified by the variance, the standard deviation, or the coefficient of variation. This can then be averaged over the distribution of true values to obtain an ensemble variance. Precision profile, which is the precision over the range of true values, could also be reported[93,94].
c. Finally, a summary FoM that quantifies the overall reliability while accounting for the variability in the parameter to be estimated, is the ensemble mean square error (EMSE). The EMSE is the mean square error averaged over the distribution of the true values and over multiple image acquisitions. The EMSE incorporates the effects of both bias and variance.

The choice of the FoM should depend on the application. For example, consider a PET-derived quantitative feature being evaluated as a biomarker to separate two patient populations. For this feature, similar bias in the estimated feature for the two populations may not be concerning. Instead, the precision of the estimated feature, which affects the separability between the two populations, would be more relevant. As another example, suppose we are interested in the change in a quantitative PET-derived feature over time; assuming constant bias over time, it is the precision of the measurements that determines when a true change over time can be discerned from just measurement noise[94,95].

**Clinical No-Gold-Standard Evaluation**: The above FoMs are applicable when the true values of the parameter of interest are known. This is the case in simulation or physical-phantom studies. However, such ground truth is typically unavailable during clinical evaluation of quantitative PET methods. To address this issue in the context of evaluating different quantitative imaging methods, no-gold-standard evaluation (NGSE) techniques have been developed[96-103], including for PET[99,104]. In these techniques, the key idea is that even though the ground truth corresponding to a measured value is unknown, since the measured value results from a specific image-formation and quantification process, a relationship between the true and measured values is expected. If the relationship is linear, it can be modeled using slope, bias, and standard deviation of a normally distributed noise term. Even in the absence of the ground truth, if we can assume that the ground truth is sampled from a bounded distribution, then we can statistically estimate these terms for the different methods using ML-based techniques. For a linear relationship between the true and measured quantitative values, the ratio of the noise standard deviation and slope terms can then be used as an FoM to evaluate the different methods on the basis of precision of the measured values[96,99,100].

### B.4.3 Joint classification and quantification task

The estimation ROC (EROC) curve provides a comprehensive description of performance on joint detection and estimation tasks[105]. To define the EROC curve, we first define a utility function that represents the closeness between the estimated and true signal parameters. The EROC is generated by plotting the expected value of the utility function in true positive cases with respect to the FPF. The area under the estimation ROC (AEROC) curve provides a summary FoM to evaluate task performance.

This concludes the description of the OAIQ framework from evaluating AI-based medical imaging methods. The framework is summarized in Fig. 5. A set of key factors to consider when designing OAIQ studies for AI methods is summarized in Table 2.

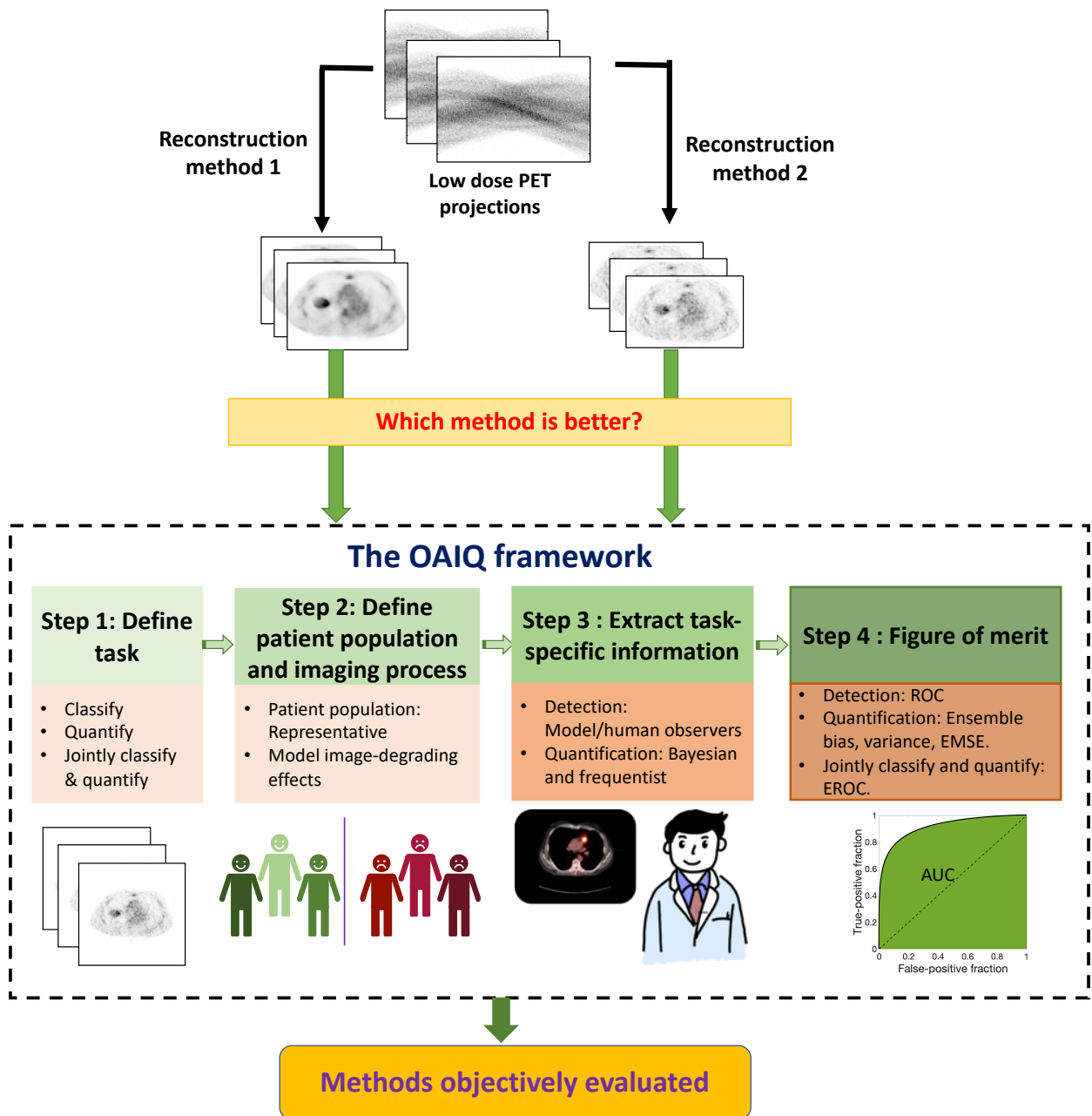

Fig. 5: A schematic of the OAIQ-based framework to evaluate AI-based methods for PET, demonstrated in the context of evaluating two low-dose reconstruction methods.

## C. Task-based evaluation of AI methods: Role of physicians

Physicians, including radiologists, nuclear medicine physicians and disease specialists, have a very important role in OAIQ-based evaluation studies of AI methods. Their inputs are vital to designing clinically relevant study designs, and their participation substantially improves the rigor of the evaluation study. They have a role in each of the four steps of the OAIQ-based framework, as summarized in Fig. 6 and described below:

a. **Step 1: Defining the task**: Physicians can help define the clinically most relevant task for the evaluation of the AI method. For example, consider an AI-based oncological PET reconstruction method. The physicians can specify, from a clinical perspective, whether the method should be evaluated on just the tumor-detection task or whether the task should include quantification of some parameter such as tracer uptake and MTV. Alternatively, the physician may specify that the goal of quantification is to use the parameter as a prognostic marker, in which case, this becomes a classification task.
b. **Step 2: Choosing patient population**: Physicians can help define patient populations that are representative and diverse as in clinical practice. They can help enrich the dataset with rare cases when those cases are absent. Additionally, in simulation studies, physicians can help define the physical and statistical properties of the simulated population.
In simulation-based evaluations, validating that the simulated and real images have similar distributions is important for clinically relevant evaluations. Observer studies with physicians can quantitatively evaluate the clinical realism of simulations. For example, Liu et al showed that a two alternative forced choice (2-AFC) study with trained nuclear-medicine physicians may help evaluate similarity of distributions of simulated and clinical populations[46].
c. **Step 3: Extract task-specific information**: As mentioned above, for a detection task conducted on reconstructed images, it is preferable if physicians can serve as observers. This is true regardless of whether the images are simulated or real. Similarly, for quantitative tasks, physicians can assist in the intermediate steps. For example, consider a PET reconstruction method being evaluated on the task of quantifying regional uptake. In this study, physicians can delineate the boundaries of the different regions.
d. **Step 4: Figures of merit**: Physicians can help define the most clinically relevant FoMs. For example, ROC analysis is a comprehensive description of performance on detection tasks. However, in clinical practice, the AI method will operate at one point on the ROC curve with a certain sensitivity and specificity. Depending on the clinical application, sensitivity or specificity may be more important. Inputs from physicians can help choose these specificity and sensitivity values for the clinical application.

   Further, when conducting clinical evaluation, the ground truth is typically not known. Sometimes multiple reference standards are needed in the same study. For example, for studies of breast-cancer detection, biopsy might be used for subjects with suspicious lesions but one-year follow-up would be the reference standard for those subjects without detected suspicious lesions. In other applications where no reference standards are available, a panel of 3-5 expert radiologists might serve as truthers, where each truther independently interprets the sample of images and makes the relevant measurements, blinded to the results of the AI algorithm and the other truthers. Then the results of the truthers are averaged, using *a priori* defined rules, to create the reference standard. Reference standards from panelists may have uncertainties associated with them. Approaches have been proposed to estimate this uncertainty[106]. In fact, the approach proposed by Miller et al[106] was used in the evaluation of the first FDA-approved computer-aided diagnosis algorithm for lung nodule detection.

Imaging scientists developing and evaluating AI-based solutions for medical imaging should closely collaborate with radiologists right from the study-design stage to ensure a clinically relevant and rigorous evaluation of these solutions.

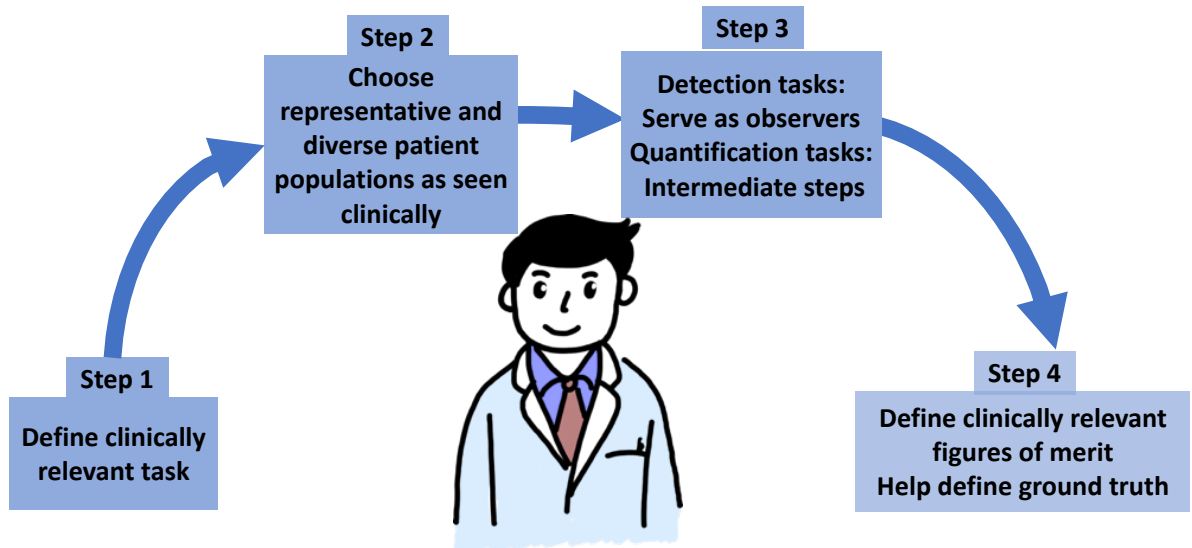

Fig. 6: A schematic showing the key role of physicians in task-based evaluation of AI methods

## D. Example Evaluation Studies

In this section, we present examples of using the described OAIQ-based framework to evaluate hypothetical AI-based methods for two applications in PET: Data acquisition and image segmentation. Both these examples are based on using realistic simulation studies. Readers are referred to recent studies using the above-described OAIQ framework to objectively evaluate AI-based methods for two other important PET applications, namely reconstruction[54] and denoising[31].

### D.1 Evaluation of a hypothetical AI-based approach for data acquisition

AI-based methods for improving PET system instrumentation are showing much promise. For example, a CNN-based approach to estimate time-of-flight (TOF) from PET-detector waveforms yielded improved performance compared to existing methods[107]. Studies have shown that incorporating TOF information helps improve lesion detection, system resolution and tumor contrast [108,109]. In this section, we demonstrate the application of OAIQ to evaluate a hypothetical AI-based method for improving timing resolution in a TOF-PET system.

*Definition of task:* The task is posed as detecting cardiac perfusion defects using myocardial perfusion PET, where the defect as well as the anatomy and physiological uptake varies across the patient population. This leads to an SKS/BKS task.

*Defining patient population and imaging process:* Realistic simulations provide a feasible mechanism to conduct this study. Recommendations outlined earlier and described in Table 2 should be followed.

*Observer:* Since the goal is system optimization, as discussed in Sec. 2.3.1, an optimal observer, such as the ideal observer or optimal linear observers, should be used.

*FoM:* ROC analysis and AUC.

*Claim from the evaluation study*: On the task of defect detection in myocardial perfusion PET for a signal-known-statistically/background-known-statistically task, an AI-based approach to data acquisition yielded an increase in AUC from X (obtained without assistance of the AI) to Y, with a change of $\Delta Z$ (95% confidence intervals), as evaluated using a realistic simulation study with a channelized linear discriminant[65].

## D.2 Evaluation of a hypothetical AI-based tumor-segmentation method

AI-based segmentation methods are showing significant promise in oncological PET, where they are being explored for tasks such as radiotherapy planning and estimating quantitative features, such as MTV, TLG, and features that describe intra-tumor heterogeneity and shape. We consider the evaluation of a hypothetical AI-based oncological PET segmentation method on the task of quantifying MTV from FDG-PET images.

*Defining the task:* Measuring MTV of the primary tumors from FDG-PET images in patients with lung cancer.

*Defining patient population and imaging process:* Since segmentations are performed on reconstructed images, for clinical realism, it is important that the reconstruction protocol be similar to those in clinical settings. Further, it is important that aspects such as intra-tumor heterogeneity and tumor shape are realistically simulated.

*Observer:* To estimate MTV for an OAIQ study, we recognize that tumors are continuous regions, while PET images are voxelized. Voxels in reconstructed images may contain a mixture of tumor and normal tissue. Thus, to estimate MTV, we should preferably measure the volume that tumor occupies in each voxel[79]. MTV can then be computed as the sum of tumor-fraction volumes within each voxel, multiplied by the volume of the voxel. However, if the segmentation method instead classifies each voxel as either tumor or background, then the MTV can be defined as the product of the number of tumor voxels and the volume of each voxel.

*FoM:* MTV is being explored as a biomarker for separating patient populations into categories, for example identifying patients likely to respond to therapy vs. those who don't. Thus, if the bias is observed to be the same across the two populations, then standard deviation would be the relevant FoM.

*Claim from the evaluation study*: An AI-based tumor segmentation method yielded MTV values that had an average normalized standard deviation of X (95% confidence intervals) in patients with stage III non-small cell lung cancer, as evaluated using a realistic simulation study.

# E. Discussions and Summary

The OAIQ framework provides a comprehensive and rigorous framework to objectively evaluate imaging systems and methods on performance in clinical tasks. Given the unique training-based nature of neural network-based AI methods, in this work we have outlined a fine-tuned OAIQ framework to evaluate these methods on performance in clinical tasks. In the process, we have highlighted factors that should be considered while designing the OAIQ study. Further, tools have been summarized to implement the concepts of this framework. Key factors to consider when designing OAIQ studies for AI-based methods have been summarized in Table 2.

The OAIQ-based evaluation strategy quantifies method performance over populations, rather than a single individual. This, in addition to being clinically realistic and comprehensive, also allows the evaluation to study aspects such as generalizability of the approach to different patient populations and acquisition protocols. For example, if there is a method that claims to be generalizable across different age groups, Step 2 of the OAIQ framework would require the method to include these different age groups in their patient population description and use these to define the distribution of the patient images while extracting the task-specific information. The OAIQ framework can also evaluate methods for specific subsets of the population if the method is only suitable for that subset. Thus, the framework can help generate appropriate claims on generalizability of the method. More generally, an OAIQ-based evaluation provides inputs to specify a descriptive and statistically rigorous claim for a method.

Table 2: A list of key factors to consider when designing task-based evaluation studies for AI methods

| OAIQ study step | Factors to consider during study design ||
|---|---|---|
| | **Task-based clinical evaluation** | **Additional considerations for evaluation with simulation studies** |
| Definition of task | Consider clinically relevant tasks | Consider tasks that model defect and background variability in patient populations: SKS or SKEV tasks |
| Choosing patient populations | Should be representative to that observed in clinical practice. Should be reported. | Anthropomorphic phantoms that are realistic and representative. Phantom resolution should be higher than reconstructed image resolution. |
| Defining the imaging process | Image acquisition protocol should be clearly reported. | Model clinical scanners realistically. Follow clinical protocols. |
| Extract task-specific information: Detection tasks | Preferably human observers | Use optimal observers operating on projection data for system optimization and anthropomorphic observers for applications that use reconstructed images. |
| Extract task-specific information: Quantification tasks | Estimate object properties, not image properties. Consider estimating properties directly from projection data. Consider estimability. ||
| Figure of merit: Detection tasks | Sensitivity, specificity, ROC analysis and AUC values. In clinical studies, define ground truth through consensus reads ||
| Figures of merit: Quantification tasks | Consider population-based metrics, such as ensemble bias and variance (or bias and variance profiles), and EMSE, to quantify accuracy, precision and overall reliability of quantification. ||

A focus in this paper has been providing tools to conduct OAIQ-based evaluation *in silico* using a VCT framework, where the patient population, imaging system and human observer are replaced by anthropomorphic phantoms, simulated systems, and model observers, respectively. One reason for the limited usage of task-based evaluation of imaging methods has been the perception that they always require expensive and time-consuming clinical studies. A reason that task-agnostic FoMs such as RMSE and Dice scores are widely used in medical imaging is that they can be evaluated without requiring such studies. However, we note that OAIQ using a VCT framework can be performed relatively quickly, inexpensively and entirely *in silico* without having to address the logistical challenges associated with clinical studies. These *in silico* studies can help identify promising candidate methods for further clinical evaluation. We hope that the description provided in this paper will stimulate wider task-based evaluation of AI methods using VCTs.

There are multiple areas of future research to improve OAIQ-based evaluation techniques. An important area is developing OAIQ-based methods for quantification tasks. Multiple important topics require investigation in this area. To begin with, current quantification methods operate on reconstructed images. However, it is recognized that such methods suffer from reconstruction-related information loss and have limited efficacy in modeling noise. Additionally, methods to quantify regional radiotracer uptake directly from projection data[38,75,77] have shown improved performance compared to reconstruction-based quantification. This motivates efforts to estimate other object parameters directly from projection data. Also, in estimation tasks, there is an important need to develop clinically relevant FoMs. While FoMs such as bias, variance and EMSE quantify the error between true and measured values, FoMs that quantify the impact of this error in making clinical decisions are needed. For example, if the objective using a quantitative method is separating patient populations into two categories, then that becomes a two-class classification task, and FoMs that quantify performance on this task, such as ROC analysis, may be an appropriate approach to study such a method. Another example is to focus on FoMs in the context of prediction: for example, evaluating and optimizing AI-methods to maximize the hazard ratio (HR) in Cox regression analysis; i.e. best stratifying patients in terms of their outcome. Another area of research is developing observers for joint classification/quantification tasks. Clinically realistic estimation tasks typically have a detection step, so developing observers for these tasks would make the evaluation study even more clinically relevant.

Clinical task-based evaluation of AI-based methods is another exciting research frontier. Current clinical evaluations of AI methods are typically restricted to some measure of "goodness" of the image, as subjectively defined through closeness to some reference standard, noise properties, or ease of interpretability of the image. An OAIQ-based evaluation paradigm should instead focus on evaluation on the clinical task. This is an area where close collaborations between physicians and AI developers is essential. In this context, Rubin provides descriptive guidelines for the specific ways through which radiologists can get involved in evaluation of AI tools and other aspects of AI method development[110]. The Society of Nuclear Medicine and Molecular Imaging AI taskforce is also preparing a set of best practices for evaluation of AI methods[33]. This will include methods for clinical task-based evaluation.

A key issue confronting the AI community is the inability of AI methods to generalize well to new unseen data. Models trained with one group of patients (say with a certain ethnicity) may underperform, or even worse, provide misleading results with other groups. The lack of interpretability in AI compounds this issue. Thus, when conducting OAIQ studies to evaluate AI methods, it is important that the patient populations chosen are representative and diverse as in clinical practice. This principle should also be followed when conducting OAIQ studies using VCTs. Here, an area that requires attention is developing techniques to ensure that the distributions of the digital phantom populations match those in clinical settings. In other words, it is not just important that the patient anatomies and physiologies are realistic for each patient, but that the distribution of these properties should also be realistic and representative.

In summary, the elucidated OAIQ-based framework presents a rigorous and comprehensive paradigm to evaluate the emerging AI-based imaging approaches on clinically relevant tasks. Such evaluation will increase the confidence in clinical utility of these methods and thus strengthen the potential for clinical translation. Our vision is that OAIQ-based VCT studies will help identify promising AI-based methods for further clinical evaluation and then OAIQ-based clinical studies will help evaluate these methods on clinical tasks. This will provide trust in the clinical application of these methods, ultimately leading to improvements in quality of healthcare. Software to conduct task-based image-quality evaluation studies is available at multiple locations such as the University of Arizona Image quality toolbox and the Metz ROC software at the University of Chicago.

## F. Acknowledgements

Financial support for this work was provided by the National Institute of Biomedical Imaging and Bioengineering Trailblazer award (R21 EB024647) and R01 EB031051. The authors thank Richard Wahl, MD and Prabhat K.C., PhD for several helpful discussions and the anonymous reviewers for their review.